\documentstyle[preprint,eqsecnum,aps]{revtex}

\begin{document}
\draft

\newcommand{\beq}{\begin{equation}}
\newcommand{\eeq}{\end{equation}}
\newcommand{\ba}{\begin{eqnarray}}
\newcommand{\ea}{\end{eqnarray}}
\newcommand{\boldsigma}{\mbox{\boldmath$\sigma$}}
\newcommand{\boldtau}{\mbox{\boldmath$\tau$}}
\newcommand{\boldomega}{\mbox{\boldmath$\omega$}}
\newcommand{\boldtheta}{\mbox{\boldmath$\theta$}}
\newcommand{\boldnabla}{\mbox{\boldmath$\nabla$}}

\title{Effects of Nonlocal One-Pion-Exchange Potential in Deuteron}
\author{J. L. Forest\cite{jlf}}
\address{Jefferson Lab Theory Group, 12000 Jefferson Ave., Newport News,
VA 23606}

\date{\today}

\maketitle

\begin{abstract}  
The off-shell aspects of the one-pion-exchange potential (OPEP)
are discussed. Relativistic Hamiltonians containing relativistic
kinetic energy, relativistic OPEP with various off-shell behaviors
and Argonne $v_{18}$ short-range parameterization are used to study
the deuteron properties. The OPEP off-shell behaviors depend on whether
a pseudovector or pseudoscalar pion-nucleon coupling is used and are
characterized by a parameter $\mu$. We study potentials having
$\mu$=$-$1, 0 and +1 and we find that they are nearly unitarily equivalent.
We also find that a nonrelativistic
Hamiltonian containing local potentials and nonrelativistic kinetic
energy provides a good approximation to a Hamiltonian containing
a relativistic OPEP based on pseudovector pion-nucleon coupling and
relativistic kinetic energy.
\end{abstract}
\vspace{.5in}
\pacs{PACS numbers: 21.30.-x, 24.10.Jv}

\newpage

\section{Introduction}

The modern high-quality two-nucleon potential models include three local
potentials: Nijmegen II~\cite{SKTS94}, Reid93~\cite{SKTS94} and Argonne
$v_{18}$ (AV18)~\cite{Wiringa95}, and two nonlocal potential:
Nijmegen I~\cite{SKTS94} and CD-Bonn~\cite{Machleidt96}.
These potentials accurately fit the $NN$ scattering data of the Nijmegen
database~\cite{Stoks93} and the deuteron binding energy, and are essentially
phase-equivalent. The three local
potentials predict very similar deuteron $D$-state probability
$P_D$ (5.70, 5.64 and 5.76\%) and triton energy
($-$7.63, $-$7.62 and $-$7.61 MeV) respectively, however the nonlocal CD-Bonn
potential gives rather different results: 4.83\% and $-$8.00 MeV.
The Nijmegen I predicts similar $P_D$ as the local Nijmegen II but slightly
larger triton energy of $-$7.72 MeV. The experimental value
of triton energy is $-$8.48 MeV, and $P_D$ is not an observable.
One may naturally ask: why is CD-Bonn so unique among the modern potentials?
Is one model better than the other, or are all modern potentials equally
good in predicting properties of nuclei after correctly including all other
ingredients such as three-body forces, relativistic effects, meson-exchange
currents, etc.?  What role do nonlocalities play in properties of light
nuclei and nuclear matter?

To answer these questions, one has to look into the microscopic description
of nuclear interactions.  Theoretically as well as experimentally it is
known that the one-pion-exchange potential (OPEP) provides the long
range part of the two-body potential $v_{NN}$. The intermediate- and
short-range parts of $v_{NN}$ are obtained by
fitting the experimental data. The three local potentials have
the same nonrelativistic on-shell OPEP at long range, but different
phenomenological parameterizations at intermediate and short range.
Nijmegen I has nonlocalities in the central part of the potential at
short range.  CD-Bonn, however, contains off-shell nonlocalities in OPEP
predicted by pseudoscalar pion-nucleon coupling as well as nonlocalities
in the phenomenological shorter range parts.  The fact that the local potentials
and the Nijmegen I model predict very similar deuteron properties
which are very different from those predicted by CD-Bonn seems to suggest that
differences in OPEP could be the main cause.

Inspired by the above comparison
among the modern potentials, we want to understand various aspects
of OPEP off-shell nonlocalities in this work.
The nonlocality in the shorter-range part of the potential may also play some
role, but it is not our primary interest here.

There has always been some ambiguity about the off-shell effects in the
two-body potentials.  Friar~\cite{Friar99} recently categorized these
ambiguities into three types: (i) those caused by an energy-dependent
potential which occur naturally when expanding energy denominators in
Schr\"odinger perturbation theory; (ii) those arising from unitary 
transformations of field variables used in the Lagrangians;
(iii) those due to different choices of relativistic Hamiltonians.
In this work we are only interested in understanding
the second type of ambiguities.
In Friar's notation, potentials having different off-shell
forms are characterized by parameters $\mu$ and $\nu$,
where the $\mu$-dependence comes from whether using pseudoscalar (PS)
or pseudovector (PV) relativistic pion-nucleon interactions,
while the $\nu$-dependence comes from the 
retardation effects. Here we are only interested in the $\mu$-dependent
off-shell behaviors, but neglect the retardation effects which seem
to be relatively unimportant. In Friar's notation, we choose $\nu$=$1/2$
(no retardation) throughout this work.

The relativistic OPEP are identical on-shell, regardless of the assumed
coupling (PS or PV), but they differ off-shell. There is no
unique description for the off-shell behavior of relativistic OPEP.
It depends upon a parameter $\mu$ whose common
choices~\cite{Friar99} are $\mu$=$-1$ (PS coupling) used in
CD-Bonn~\cite{Machleidt96}, $\mu$=$0$ (minimal nonlocality) used in our
earlier work~\cite{Forest99}, and $\mu$=+1 (PV coupling) favored by
conventional Chiral Perturbation Theory (CPT).
As suggested by Friar~\cite{Friar80,Friar77} two decades ago,
two-body potentials differing in the value of $\mu$ are related by unitary
transformations up to order $1/m^2$. All these potentials
are correct to this order, and predict identical observables, even though
they have different forms.  In this view, when various
two-body potentials are combined with their corresponding three-body
potentials, the triton energy should be accurately predicted;
when electromagnetic current operators are treated consistently
with the potential, the electromagnetic observables
should be independent of the choice of $\mu$~\cite{Coon86}.
This is rather interesting
and useful because if there indeed exists such unitary equivalence,
then we could choose the simpler and computationally-easier local potentials
together with their three-body forces, and consistent electromagnetic charge
and current operators to study properties of light nuclei.

The purpose of this paper is to study deuteron properties with
potentials containing $\mu$=$-$1, 0, +1 relativistic OPEP (Sec. II),
and to examine the unitary equivalence of these potentials (Sec III).
We also attempt to find local potentials which
would give a smaller value for deuteron $P_D$, characteristic of the
$\mu$=$-1$ nonlocal potentials, but with no success (Sec IV).
The conclusions are given in Sec. V.  Some of the detailed derivations
involved in this work are given in an Appendix.

\section{Off-shell effects in deuteron}

Consider a relativistic Hamiltonian
\beq
H_R = 2\sqrt{p^2+m^2}-2m + v_{\pi} + v_R,
\label{HR}
\eeq
in the frame in which the deuteron is at rest. The transformation of this
$H_R$ to other frames is discussed in Ref. \cite{Forest95}.
Here $v_{\pi}$ is the OPEP containing off-shell nonlocalities, and $v_R$ is
the remaining part of the potential which is phenomenological. In this work
we use AV18 parameterization for $v_R$.
As suggested by Friar~\cite{Friar99}, after neglecting retardation effects,
the relativistic OPEP can be expressed in the following general form
\ba
v_{\pi}^{\mu}({\bf p}^{\prime},{\bf p}) &=&
- \frac{f^2_{\pi NN}}{m_{\pi}^2}\ \frac{\boldtau_i \cdot \boldtau_j}
{m_{\pi}^2+q^2}\ \frac{m}{E}\ \frac{m}{E^{\prime}}\ 
\left[ \frac{}{}\boldsigma_i \cdot {\bf q}\ \boldsigma_j \cdot {\bf q} \right.
\nonumber \\
& & \left. +\mu\times (E^{\prime}-E) \left( \frac 
{\boldsigma_i \cdot {\bf p}^{\prime}\ \boldsigma_j \cdot {\bf p}^{\prime}}
{E^{\prime}+m} - \frac{\boldsigma_i \cdot {\bf p}\ \boldsigma_j \cdot {\bf p}}
{E+m} \right) \right].
\label{vopep}
\ea
Here $m_{\pi}$ and $m$ are the pion and nucleon mass respectively,
$f_{\pi NN}$ is the pion-nucleon coupling constant, ${\bf p}$ and
${\bf p}^{\prime}$ are the momenta of particle $i$ in the center of mass
frame before and after the 
interaction, ${\bf q}$=${\bf p}^{\prime}-{\bf p}$ is the momentum transfer,
$E$=$\sqrt{m^2+p^2}$, and $E^{\prime}$=$\sqrt{m^2+p^{\prime 2}}$.
The $\mu$-dependent term corresponds to nonlocalities in configuration space;
it vanishes on-shell where $E$=$E^{\prime}$.
When used in momentum space to solve for deuteron properties,
equation (\ref{vopep}) is multiplied by the $\pi NN$ form factor $F(q)$
to ensure convergence
\beq
F(q) = \left(\frac{\Lambda_{\pi}^2-m_{\pi}^2}{\Lambda_{\pi}^2+q^2}\right)^2.
\label{ff}
\eeq
In the present work we use the cutoff mass $\Lambda_{\pi}$=1.2 GeV.

The expression for $\mu$=$-1$, i.e., $v_{\pi}^{\mu=-1}$ can be easily derived
using second-order covariant perturbation theory with a pseudoscalar
pion-nucleon coupling
\beq
H_{\pi NN}^{\rm PS} = iG\bar{\psi}\gamma_5\tau_i\psi\phi_i
\eeq
with $G^2$=$4m^2f_{\pi NN}^2/m_{\pi}^2$.
When applying the same technique to the pseudovector-coupling
\beq
H_{\pi NN}^{\rm PV} = -\frac{f_{\pi NN}}{m_{\pi}}\bar{\psi}\gamma^{\mu}\gamma_5
\tau_i\psi\partial_{\mu}\phi_i,
\eeq
we can not obtain the $\mu$=$+1$ expression $v_{\pi}^{\mu=+1}$. The reason is that
the derivative coupling of the pion field yields a term dependent on the 
pion energy $q_0$, and within second-order diagrams conservation of energy
at each vertex requires $q_0$=0. The expression corresponding to $q_0$=0
has been given in Ref. \cite{Adam93} up to order $p^2/m^2$, and is different
from $v_{\pi}^{\mu=+1}$ of that order. The correct expression
for $v_{\pi}^{\mu=+1}$ [Eq. (\ref{vopep})] can be obtained from fourth-order
calculations~\cite{Sugawara60}.

The $\mu$=0 (``minimal nonlocality'') corresponds to on-shell relativistic OPEP
\beq
v_{\pi}^{\mu=0}({\bf p}^{\prime},{\bf p}) = - \frac{f^2_{\pi NN}}{m_{\pi}^2}\ 
\frac{\boldsigma_i \cdot {\bf q}\ \boldsigma_j \cdot {\bf q}
\boldtau_i \cdot \boldtau_j}{m_{\pi}^2+q^2}
\ \frac{m}{E}\ \frac{m}{E^{\prime}},
\label{vrel}
\eeq
and has been studied in detail in Ref. \cite{Forest99}.
In the nonrelativistic limit, $E\approx E^{\prime}\approx m$,
and Eq. (\ref{vrel}) yields nonrelativistic local OPEP
used in the three local potentials (Nijmegen II, Reid93 and AV18)
\beq
v_{\pi}^{\rm NR}({\bf q}) = - \frac{f^2_{\pi NN}}{m_{\pi}^2}\ 
\frac{\boldsigma_i \cdot {\bf q}\ \boldsigma_j \cdot {\bf q}
\boldtau_i \cdot \boldtau_j}{m_{\pi}^2+q^2}.
\label{vnr}
\eeq

When Fourier transformed to configuration space, $v_{\pi}^{\rm NR}$
[Eq. (\ref{vnr})] becomes the familiar sum of the tensor and spin-spin
interaction terms. The $\delta$-function singularity present in the
spin-spin term is usually removed 
or ``smeared'' by various form factors, e.g., exponential form factor
in Nijmegen II and AV18, dipole form factor in Reid93 and
CD-Bonn [Eq. (\ref{ff})].  We are interested in understanding the difference
between the local and nonlocal potentials. It is sufficient to consider only
one of the three available representations of a local potential, which we
choose to be AV18. The results can probably be generalized to all modern
local potentials.

The OPEP in AV18 is given by the following expression in configuration space
\beq
v_{\pi}({\bf r}) = \frac{1}{3}m_{\pi} \frac{f_{\pi NN}^2}{4\pi}
\left[Y_{\pi}(r)\boldsigma_i\cdot\boldsigma_j+T_{\pi}(r)S_{ij}\right]
\boldtau_i\cdot\boldtau_j,
\label{av18}
\eeq
where
\ba
Y_{\pi}(r)&=&\frac{e^{-m_{\pi} r}}{m_{\pi} r}\left(1-e^{-cr^2}\right), \\
T_{\pi}(r)&=&\left(1+\frac{3}{m_{\pi} r}+\frac{3}{(m_{\pi} r)^2}\right)
\left(1-e^{-cr^2}\right)^2, \\
S_{ij}&=&3\boldsigma_i\cdot\hat{\bf r}\boldsigma_j\cdot\hat{\bf r}-\boldsigma_i\cdot
\boldsigma_j.
\label{more-av18}
\ea
The rest of the AV18 is phenomenological and uses the Woods-Saxon and
$T_{\pi}^2$ functions. When projected onto spin $S$=1 and isospin $T$=0
channel, the overall potential is expressed as
\beq
v_{S=1,T=0}(r) = v_c(r) + v_t(r)S_{ij} + v_{ls}(r){\bf L}\cdot{\bf S}
+ v_{l2}(r)L^2 + v_{ls2}(r)({\bf L}\cdot{\bf S})^2,
\label{v10}
\eeq
where the five terms are called the central, tensor, spin-orbit,
quadratic orbital angular momentum and quadratic spin-orbit terms,
respectively. The OPEP is obviously included in the central term and
is the main contribution to the tensor term.

It is interesting to study deuteron properties using the isoscalar part of AV18
but replace its OPEP with $v_{\pi}^{\mu=-1,0,+1}$ (for future
reference, these potentials are denoted as $v_{18}^{\mu=-1,0,+1}$). 
We construct the relativistic Hamiltonian containing $v_{18}^{\mu}$ and
relativistic kinetic energy to be phase equivalent to the nonrelativistic
Hamiltonian containing the isoscalar part of AV18 and nonrelativistic kinetic
energy. The short-range part of the potential $v_R$ depends on $\mu$
implicitly in order to reproduce the phase shifts.  We adjust the fifteen free
parameters in $v_R$ and fit the phase shifts in $S$=1, $T$=0 channel,
i.e., $^3S_1$,
$E_1$, $^3D_1$, $^3D_2$, $^3D_3$, $E_3$, $^3G_3$, and the deuteron
binding energy with great accuracy.  The deuteron properties are listed
in Table I and the deuteron wave functions are plotted in
Fig. \ref{fig:wavefunc}.
The first row of Table I shows the experimental results; it is followed
by results obtained with five modern potentials, and the last
three rows are the results obtained with $v_{18}^{\mu=-1,0,+1}$.

The $v_{18}^{\mu=0}$ has been discussed in detail in Ref. \cite{Forest99}.
It produces very similar deuteron $D$-state probability ($P_D$=5.73\%) to
that of the nonrelativistic AV18 (5.76\%). Their deuteron wave functions are
also similar as shown in Fig. \ref{fig:wavefunc}. The $v_{18}^{\mu=0}$
binds the triton slightly more ($\sim$0.1 MeV) than AV18~\cite{Forest99}.
These similarities between $v_{18}^{\mu=0}$ and AV18 seem
reasonable because the Hamiltonian is constrained
such that the nonlocalities from the $v_{18}^{\mu=0}$ and relativistic
kinetic energy $T_R$ have to cancel exactly in the deuteron to give the same
binding energy. The small variations in $P_D$ and triton binding energy
are due to the almost perfect cancellation between $v_{18}^{\mu=0}$ and
$T_R$ nonlocalities.

In the cases of $\mu$=$-1$ and +1, there does not seem to have such a perfect
cancellation.
The deuteron $P_D$ predicted by $v_{18}^{\mu=-1}$ is 4.98\%, noticeably
smaller than that of $v_{18}^{\mu=0}$ (5.73\%), while the $P_D$ of
$v_{18}^{\mu=+1}$ is 6.26\%, noticeably larger than that of $v_{18}^{\mu=0}$.
This is entirely due to the off-shell terms in $v_{\pi}^{\mu=-1}$
and $v_{\pi}^{\mu=+1}$ absent in $v_{\pi}^{\mu=0}$,
because $v_{18}^{\mu=-1,0,+1}$ all have the same short-range parameterization of
AV18; only the values of parameters are slightly different in order to
fit the data.
The $P_D$ of $v_{18}^{\mu=-1}$ (4.98\%) is very close to that of CD-Bonn
(4.83\%) as expected since both potentials have the same OPEP
$v_{\pi}^{\mu=-1}$ apart from different pion cut-off mass used in the form
factor [Eq. (\ref{ff})]. We use 1.2 GeV for $\Lambda_{\pi}$ while CD-Bonn
uses 1.7 GeV. When we use $\Lambda_{\pi}$=1.7 GeV in our $v_{\pi}^{\mu=-1}$
as a test and refit the phase shifts, the deuteron properties remain almost
the same ($P_D$=4.97\%, $D/S$ ratio=0.0240, $Q_d$=0.0267 fm$^2$). Therefore
the differences between $v_{18}^{\mu=-1}$ and CD-Bonn must come
from those in the short-range parts.
This is further confirmed by the similar $D$-state wave function at
long range ($r>1.5$ fm).  The difference in the deuteron quadrupole moment
given by the two potentials is $\sim$1.5\%. It is smaller than the 3\%
difference in $P_D$. This can be understood from the following arguments.
In the impulse approximation, deuteron $Q_d$ is given by
\beq
Q_d = \frac{1}{\sqrt{50}}\int_{0}^{\infty}(u(r)w(r)-w^2(r)/\sqrt{8})r^2dr,
\eeq
and it is an ``outside'' quantity whose major contributions come from
the longer-range part of the wave functions which is fairly model-independent.
On the other hand, the $D$-state probability $P_D$ defined as
\beq
P_D=\int_{0}^{\infty}w^2(r)dr,
\eeq
is an ``inside'' quantity lacking the extra factor $r^2$ in the
integrand. Therefore $P_D$ is sensitive to the interior region of the
wave functions where larger differences exist among various potential models.

Figure \ref{fig:wavefunc} also shows another interesting point.  The
AV18 wave functions are in-between those of $v_{18}^{\mu=+1}$ and
$v_{18}^{\mu=0}$. The $v_{18}^{\mu=-1}$ wave functions are farther away,
and those of CD-Bonn are the farthest.  The difference between
the wave functions of AV18 and $v_{18}^{\mu=+1}$ favored
by CPT is rather small.
This can be easily understood from the following first-order perturbative
estimates. 
The largest contributions of OPEP come from coupling states with small $p$ to
large $p^{\prime}$. Therefore we consider the extreme case where $p$=0,
${\bf p}^{\prime}$=${\bf q}$, and expand Eq. (\ref{vopep}) in powers
of $q^2/m^2$. This gives
\ba
v_{\pi}^{\mu=-1}({\bf q},0) &=& v_{\pi}^{\rm NR}\,\left(1-\frac{3q^2}{4m^2}
+\cdots\right), \\
v_{\pi}^{\mu=0}({\bf q},0) &=& v_{\pi}^{\rm NR}\,\left(1-\frac{q^2}{2m^2}
+\cdots\right), \\
v_{\pi}^{\mu=+1}({\bf q},0) &=& v_{\pi}^{\rm NR}\,\left(1-\frac{q^2}{4m^2}
+\cdots\right).
\ea
If we treat OPEP as a perturbative term in the Hamiltonian,
the first order correction to the amplitude of a state
with large momentum ${\bf q}$ is given by
\beq
\psi({\bf q}) \approx \frac{v_{\pi}({\bf q},0)}{E(q)-E(0)}\psi({\bf q}=0).
\eeq
Here $v_{\pi}$ can be $v_{\pi}^{\rm NR}$ or $v_{\pi}^{\mu}$ depending on
whether the Hamiltonian is nonrelativistic or relativistic.
In the nonrelativistic case, $E(q)-E(0)\approx q^2/m$, while in the
relativistic case, $E(q)-E(0)\approx q^2/m\times (1-q^2/4m^2)$. Therefore
the difference between the first order, large ${\bf q}$,  amplitudes
of $v_{\pi}^{\mu}$ and
$v_{\pi}^{\rm NR}$ is given by
\ba
\psi^{\mu=-1}({\bf q}) - \psi^{\rm NR}({\bf q}) &\approx&
-\frac{q^2}{2m^2}\psi^{\rm NR}({\bf q}) \\
\psi^{\mu=0}({\bf q}) - \psi^{\rm NR}({\bf q}) &\approx&
-\frac{q^2}{4m^2}\psi^{\rm NR}({\bf q}) \\
\psi^{\mu=+1}({\bf q}) - \psi^{\rm NR}({\bf q}) &\approx& 0.
\ea
This estimate suggests that the $v_{\pi}^{\mu=+1}$ wave function is
the closest to the nonrelativistic wave function, that of $v_{\pi}^{\mu=0}$
is the second closest, and that of $v_{\pi}^{\mu=-1}$ is the farthest.
In reality, other components of $v({\bf p}^{\prime},{\bf p})$
contribute as well, and the OPEP can not be treated perturbatively. The exact
calculations show (Fig. \ref{fig:wavefunc}) that the nonrelativistic wave
function is in-between those of $v_{18}^{\mu=+1}$ and $v_{18}^{\mu=0}$.
This implies that the wave functions obtained from nonrelativistic local
potentials such as AV18 and nonrelativistic kinetic energy provide excellent
approximation to those with $v_{18}^{\mu=+1}$ and $v_{18}^{\mu=0}$ and
relativistic kinetic energy.

The expectation values of the kinetic energy $T$, OPEP $v_{\pi}$ and
the phenomenological short-range part $v_R$ of various potentials are
compared in Table II. As we can see, the individual terms change slightly
among various potentials, but the sum of them gives identical total energy
of $-$2.242 MeV by construction. The experimental value of deuteron binding
energy ($-$2.224 MeV) is obtained from the full AV18 with electromagnetic
interactions. From Table II, it appears that the individual terms of
$v_{18}^{\mu=+1}$ are the closest to those of AV18, those of $v_{18}^{\mu=0}$
are the second closest, and those of $v_{18}^{\mu=-1}$ are the farthest.
Also listed in this table are the on-shell ($v_{\pi}^{\rm on-shell}$)
and off-shell ($v_{\pi}^{\rm off-shell}$) contributions to the OPEP.
The sum of the two terms gives the total $v_{\pi}$ [Eq. (\ref{vopep})].
For AV18 and $v_{18}^{\mu=0}$, the off-shell contributions are obviously
zero, and the on-shell contributions are the nonrelativistic
OPEP [Eq. (\ref{av18})] and on-shell relativistic OPEP [Eq. (\ref{vrel})],
respectively. For $v_{18}^{\mu=\pm 1}$, the off-shell contribution is much
smaller than the on-shell one because the former is of order $p^2/m^2$ of
the latter in the leading order.

The deuteron charge form factors $F_c^2$ and $T_{20}$ calculated in impulse
approximation are plotted in Figs. \ref{fig:fc} and \ref{fig:t20}.  
Notice that the results predicted by AV18 are in-between
those of $v_{18}^{\mu=+1}$ and $v_{18}^{\mu=0}$. 
The difference between the $T_{20}$ obtained with CD-Bonn and AV18 is much
larger than that calculated with $v_{18}^{\mu}$ of various choices of $\mu$
implying that the nonlocality in the shorter-range part of CD-Bonn
is the main cause.

\section{Unitary equivalence}

Friar~\cite{Friar80,Friar77} suggested two decades ago that the physics
of pseudovector and pseudoscalar couplings of pions to nucleons is related
via the equivalence theorem, originating from the Dyson
transformation~\cite{Dyson48}. Basically, the theorem states~\cite{Friar77}
that the one-pion-exchange currents, together with matrix elements of
the impulse charge operator calculated with a potential including OPEP,
must be the same for both PS and PV couplings to order $G^2$.
As an interesting byproduct of Friar's work, the unitary transformation
involved can change the deuteron $D$-state probability, indicating that
this quantity is not measurable~\cite{Friar79}.

It is interesting to test this theorem using our relativistic Hamiltonian
[Eq. (\ref{HR})] for $\mu$=$-$1, 0, +1. The Hamiltonian also contains a
short-range potential $v_R$ which is implicitly dependent on $\mu$ in order
to fit the data.  We can rewrite Eq. (\ref{HR}) as the following
\beq
H^{\mu}_{\rm exact} = T_R + v_{\pi}^{\mu=0} + \Delta v_{\pi}^{\mu} + v_R^{\mu}
\label{Hmu}
\eeq
where $T_R$=$2\sqrt{p^2+m^2}-2m$ is the relativistic kinetic energy,
$v_{\pi}^{\mu=0}$ is the on-shell relativistic OPEP given by Eq. (\ref{vrel})
and $\Delta v_{\pi}^{\mu}$ is the off-shell term in Eq. (\ref{vopep}) given by
\beq
\Delta v_{\pi}^{\mu}({\bf p}^{\prime},{\bf p}) = -\mu\times\frac{f^2_{\pi NN}}
{m_{\pi}^2}\ \frac{\boldtau_i \cdot \boldtau_j}{m_{\pi}^2+q^2}\ \frac{m}{E}
\ \frac{m}{E^{\prime}}\ (E^{\prime}-E)
\left( \frac{\boldsigma_i \cdot {\bf p}^{\prime}\ \boldsigma_j \cdot
{\bf p}^{\prime}}{E^{\prime}+m} - 
\frac{\boldsigma_i \cdot {\bf p}\ \boldsigma_j \cdot {\bf p}}{E+m} \right).
\eeq
In light nuclei, the small binding energy comes from the
large cancellation between the kinetic energy and two-body potential.  Hence
$v_{\pi}^{\mu=0}$ and $v_R^{\mu}$ in Eq. (\ref{Hmu}) are of the same
order of magnitude as $T_R$. The off-shell term $\Delta v_{\pi}^{\mu}$
is much smaller, its leading term is of order $p^2/m^2$ of $v_{\pi}^{\mu=0}$.
It seems that in deuteron the off-shell term $\Delta v_{\pi}$ is
comparable to $v_R$ as shown in Table II. This is because deuteron
is a loosely bound state with a very large rms radius,  thus the OPEP accounts
for more than 95\% of the two-body potential, yielding a very small $v_R$.
In light nuclei with $A>2$, the OPEP still accounts for a large portion of
the two-body potential ($>$70\%), but $v_R$ is about the same order of
magnitude as the kinetic energy and OPEP~\cite{Pudliner95} and
$\Delta v_{\pi}^{\mu}$ is probably much smaller than $v_R$.

The Hamiltonian [Eq. (\ref{Hmu})] can also be expressed,
up to order $p^2/m^2$, as the following unitary transformation
\beq
H^{\mu}_{\rm uni}=e^{-iU}H_0e^{iU}\approx H_0 + i\left[H_0,U\right],
\label{Huni}
\eeq
where $H_0$ can be conveniently chosen as
\beq
H_0 = H^{\mu=0}_{\rm exact} = T_R + v_{\pi}^{\mu=0} + v_R^{\mu=0}.
\label{H0}
\eeq
The unitary operator which satisfies Eq. (\ref{Huni}) is found to be
\beq
iU = -\mu\times\frac{1}{2}\frac{f_{\pi NN}^2}{m_{\pi}^2}
\frac{\boldtau_i \cdot \boldtau_j}{m_{\pi}^2+q^2}\ \frac{m}{E}\ 
\frac{m}{E^{\prime}}\left(\frac{\boldsigma_i \cdot {\bf p}^{\prime}
\ \boldsigma_j \cdot {\bf p}^{\prime}}{E^{\prime}+m}
- \frac{\boldsigma_i \cdot {\bf p}\ \boldsigma_j \cdot {\bf p}}{E+m}\right).
\label{unitary-op}
\eeq
It can be easily verified that the commutator
\beq
\left[T_R, iU\right] = \Delta v_{\pi}^{\mu}.
\eeq
The wave function is transformed, consistently  with the Hamiltonian,
in the following way
\beq
|\psi^{\mu}\rangle^{\rm uni} = e^{-iU}|\psi_0\rangle\approx (1-iU)|\psi_0\rangle
\label{unitary-wave}
\eeq
where $|\psi_0\rangle$ is the wave function of $H_0$ obtained
by solving Schr\"{o}dinger equation. The methods used to
solve Eq. (\ref{unitary-wave}) for the deuteron are given in the Appendix.

Earlier works by Friar~\cite{Friar99,Friar80}, Desplanques and
Amghar~\cite{Desplanques92} and Adam {\it et al.}~\cite{Adam93}
used a nonrelativistic kinetic energy $p^2/m$ in the Hamiltonian, and
kept the leading order ($p^2/m^2$) terms in the ``relativistic''
OPEP and the unitary operator.  This seems to be inconsistent with the
relativistic considerations, nor is it
computationally easier than the exact treatment since all the calculations 
involving the OPEP off-shell term are performed in momentum space.
In this work we use a relativistic kinetic energy and keep the relativistic
OPEP and unitary operator $iU$ in their exact forms.

From Eqs. (\ref{Huni}), (\ref{H0}) and (\ref{unitary-op}), we find that
the unitarily transformed Hamiltonian is not exactly the same as
the $H^{\mu}_{\rm exact}$
\beq
H^{\mu}_{\rm uni} = H^{\mu}_{\rm exact} + \Delta H^{\mu}.
\eeq
The correction term is
\beq
\Delta H^{\mu} = v_R^{\mu=0} + \left[v_{\pi}^{\mu=0}, iU\right]
+ \left[v_R^{\mu=0},iU\right] - v_R^{\mu}.
\eeq
The two commutators $\left[v_{\pi}^{\mu=0}, iU\right]$ and
$\left[v_R^{\mu=0},iU\right]$ are of the same order as $\Delta v_{\pi}^{\mu}$,
but the first commutator has the range of two-pion exchange interaction while
the second one has even shorter range. Both can be absorbed into the
phenomenological short-range part
$v_R^{\mu=0}$ to obtain $v_R^{\mu\prime}$. If this $v_R^{\mu\prime}$ is
exactly the same as $v_R^{\mu}$, then $H^{\mu}_{\rm uni}$=$H^{\mu}_{\rm exact}$,
and $H^{\mu}$ is unitarily equivalent to $H^{\mu=0}$. However, it is
unlikely that $v_R^{\mu\prime}$=$v_R^{\mu}$ as demonstrated in
the wave functions shown in Fig. \ref{fig:unitary} and deuteron properties
listed in Table III.

As we can see in Fig. \ref{fig:unitary}, the unitarily transformed deuteron
wave functions are slightly different from the exact wave functions
obtained by solving the Schr\"odinger equation, mostly at short range.
The unitary transformation works rather well for the $D$-state. It also
works well for the $S$-state at long range but somewhat breaks down at very
short range ($r<0.3$ fm). This can be understood from the following arguments.
The unitary transformation discussed in this work is related to
OPEP. The long-range parts of the wave
functions are mostly determined by the OPEP, thus they are related very
well by the unitary transformation.
The $D$-state is most sensitive to the tensor force whose main contribution
comes from the OPEP. The $S$-state, however, is sensitive to both
central and tensor parts of the potential at short range, thus the unitary
transformation starts to deviate from the exact $S$-state wave function at
$r<0.3$ fm.  This deviation at such short range may have very small impact
on the observables.  For example, the deuteron binding energies calculated
with the unitarily transformed wave functions for $\mu$=$-1$ and +1 are
reduced by only $\sim$8\% as compared with the exact result, indicating that
the unitary transformation is a good approximation to the exact solution.

The deuteron properties obtained with unitary transformation are compared
with those from solving Schr\"odinger equation in Table III.  The differences
between the exact solutions and the results of unitary transformation
are obviously caused by the short-range part of the potential in $\Delta H$. 
These differences are rather small indicating that the potentials $v_{18}^{\mu}$
for $\mu$=$-$1, 0 and +1 are, to large extent, unitarily equivalent.
Desplanques and Amghar~\cite{Desplanques92} studied unitary equivalence between
Paris ($\mu=0$) and the older Bonn ($\mu$=$-1$) potentials before the birth of
the modern potentials. The differences between the results of Bonn and 
the unitarily transformed Paris, or Paris and unitarily transformed Bonn seem
to be larger than the present work, indicating that the Paris and Bonn
potentials are not as unitarily equivalent as our $v_{18}^{\mu}$ models.
The reason is probably due to the rather different structures in
the short-range part of the two potentials, and possibly due to using the
leading order terms in kinetic energy, OPEP and unitary operator
instead of their exact expressions.
Adam {\it et al.}~\cite{Adam93} did a similar study as the present work,
except that they used a pure OPEP model to describe the two-nucleon system,
and they also used leading order expressions for kinetic energy, OPEP
and unitary operator.
They studied deuteron properties for various values of $\mu$ and $\nu$.
Their deuteron $P_D$ and $Q_d$ for $\nu$=$1/2$ (no retardation) and
$\mu$=$-$1, 0, +1 seem to be larger than ours, probably because they used
a pure OPEP model instead of a realistic model which fits $NN$ phase shifts
as well as deuteron binding energy. However, the amount of changes from
before and after unitary transformation are similar to ours.

The deuteron $P_D$ of current work is plotted in Fig. \ref{fig:pd}.
The $P_D$ obtained 
with unitary transformation seems to be linear in $\mu$, while the exact
results are slightly nonlinear. This can be easily understood from the
expression of the unitarily transformed wave functions given in
Eq. (\ref{unitary-SDwave}).  We can rewrite the wave functions as
\ba
R_0^{\rm uni}(p) &=& R_0(p) + \mu\times\delta R_0(p), \nonumber \\
R_2^{\rm uni}(p) &=& R_2(p) + \mu\times\delta R_2(p),
\ea
where $R_0(p)$ and $R_2(p)$ are the normalized $S$- and $D$-state radial
wave functions, $\delta R_0(p)$ and $\delta R_2(p)$ are the second terms in
Eq. (\ref{unitary-SDwave}). In the dominant region of the wave functions,
i.e., $p<5$ fm$^{-1}$, $R_0(p)\gg \delta R_0(p)$, $R_2(p)\gg \delta R_2(p)$.
The deuteron $D$-state probability is given by
\ba
P_D^{\rm uni} &=& P_D^0 + \mu\times\left[-2P_D^0\int_0^{\infty}R_0(p)
\delta R_0(p)p^2dp+2(1-P_D^0)\int_0^{\infty}R_2(p)\delta R_2(p)
p^2dp\right] \nonumber \\
& & \hspace{0.65cm}+\ \mbox{terms of order}\ \mu^2\ \mbox{and higher},
\ea
where $P_D^0$ is the deuteron $D$-state probability of $\mu$=0.
The terms of order $\mu^2$ and higher are negligibly small and can be
neglected, therefore $P_D$ obtained from the unitary transformation
is approximately linear to $\mu$.
The exact $P_D$ is slightly nonlinear reflecting the difference of
the short-range part of the potentials.

\section{Local potentials}

Before we draw any conclusions, let's ask a question: is the small $P_D$ a
unique feature from $v_{\pi}^{\mu=-1}$, or the large $P_D$ a unique feature
from $v_{\pi}^{\mu=+1}$\,?  In other words, can we find a local potential
which fits experimental data and still gives $P_D<5$\% or $P_D>6$\%?

To explore such a possibility, we attempt to lower the $P_D$ to $\sim$5\%
by using AV18 as trial local potential in our Hamiltonian, together with
a nonrelativistic kinetic energy.
There are fifteen free parameters in the AV18 representation
[Eq. (\ref{v10})].
Just by varying these parameters to fit the data and simultaneously
constraining $P_D\approx 5\%$ can result in good fits to $^3S_1$,
$^3D_1$, $^3D_2$, $^3D_3$, $E_3$ and $^3G_3$ in addition to deuteron
binding energy, but not to the mixing angle $E_1$. At higher energies
(200--400 MeV) $E_1$ comes out significantly below the Nijmegen values.

Since $P_D$ is sensitive to the tensor force, in order to get smaller $P_D$,
we need to reduce the tensor force. We have tried various approaches
summarized below: (i) Multiply the entire tensor force $v_t(r)$ by a
factor $(1-e^{-c_1r^2})$, and adjust the parameter $c_1$ in addition to the
fifteen parameters; (ii) Add a factor $c_2r(1-e^{-c_3r^2})$ to the tensor force,
and vary the parameters $c_2$ and $c_3$ in addition to the fifteen parameters;
(iii) Use the wave functions obtained with $v_{18}^{\mu=-1}$
whose $P_D$ is 4.98\% and solve the Schr\"odinger equation for
local $v_c$ and $v_t$ (denote them as ``local $\mu$=$-1$''),
keeping $v_{ls}$, $v_{l2}$ and $v_{ls2}$
fixed. These potentials are compared with those in the AV18
in Fig. \ref{fig:vct}.
As we can see, the ``local $\mu$=$-1$'' $v_c$ and $v_t$ are both reduced 
at the short-range to reproduce the deuteron binding energy and obtain
a smaller $P_D$. We then multiply or add various functions to $v_c$ and
$v_t$ to reproduce this effect.  None of these approaches
yield a good $\chi^2$ fit.  The largest discrepancies are for $E_1$.

This is not
surprising because $E_1$ is most sensitive to the tensor force, unlike
any other states. When restricting $P_D$ to be
$\sim$5\%, we are reducing the tensor force which in turn reduces the $E_1$.
Other states can make trade-offs among $v_c$, $v_{ls}$, $\cdots$.
In contrast, when the OPEP tensor force is modified by the
off-shell term in Eq. (\ref{vopep}),
only the off-diagonal elements $v(p,p^{\prime})$ is reduced, but the 
Born term $v(p,p)$ remains the same.  This seems to suggest two things:
(i) the tensor Born term plays an important role in fitting to $E_1$ phase
shift; (ii) weakening the tensor force off-diagonally does not seem to
affect $E_1$, however it produces a smaller $P_D$ of the deuteron.

In summary, we couldn't find a local potential which gives a deuteron $P_D$
as small as 5\%. We can probably extrapolate this statement to $P_D$ as large
as 6\%. Therefore, it seems that the different deuteron $P_D$'s are associated
with the various off-shell behaviors of the OPEP.

\section{Conclusions}

We studied deuteron properties with potentials containing $\mu$=$-$1, 0, and +1
OPEP off-shell behaviors. We tested the unitary equivalence among
these potentials by comparing to the exact solutions of Schr\"odinger
equation, and we found the following interesting results:

(i)  The off-shell term involving the coupling of nucleon spin and
momentum is the primary cause for the smaller deuteron $P_D$ for
CD-Bonn and $v_{18}^{\mu=-1}$, and consequently a higher triton binding energy.
We also found that it is not possible to get a deuteron with
$P_D$ as small as 5\% with the local OPEP [Eq. (\ref{av18})].

(ii) The deuteron wave functions, charge form factors and $T_{20}$
for $\mu$=+1 favored by CPT are very close to those of the nonrelativistic AV18.

(iii) The $v_{18}^{\mu=-1,0,+1}$ models are, to large extent,
unitarily equivalent.

(iv) The differences due to unitary transformations related to OPEP
are smaller than the difference between AV18 and CD-Bonn potential.

The charge form factors and $T_{20}$ shown in this work are calculated
in impulse approximation.  After including the meson-exchange
current operator obtained with unitary transformation consistent
with the Hamiltonians and wave functions, the deuteron
$A$ and $B$ structure functions as well as $T_{20}$ turn out to be rather
similar for $v_{18}^{\mu=-1,0,+1}$ and AV18, but they differ significantly
from those of CD-Bonn~\cite{Schiavilla99}.  This suggests that AV18
and CD-Bonn are not well-related by the kind of unitary transformation 
discussed in this work, i.e., the one that deals with OPEP.

In summary, we find that a nonrelativistic Hamiltonian containing local
potentials (AV18, Nijmegen II, Reid93) and nonrelativistic kinetic energy
provides an excellent approximation to a relativistic Hamiltonian
containing a relativistic OPEP of pseudovector pion-nucleon coupling and
relativistic kinetic energy, in predicting properties of nuclei.
This may matter little for two-nucleon systems such as the deuteron discussed
in this work, or three-nucleon systems such as the triton discussed in
Refs.~\cite{Machleidt96,Friar93} for which momentum space as well as
configuration space computational methods can be applied.  However,
when we go up to nuclei having $A>3$, momentum-space techniques
face great computational difficulties, leaving the configuration-space
calculations as the main approach for high-accuracy
computations~\cite{Pudliner95,Carlson98}, in which case a local nonrelativistic
Hamiltonian is to be preferred.

\section*{Acknowledgments}
I wish to thank V. R. Pandharipande and R. Schiavilla for their valuable
input and many contributions to this subject. I have benefited
tremendously from many inspirational conversations with them.
I also thank A. Arriaga, J. L. Friar, and R. Machleidt
for many interesting discussions.
The calculations were made possible by generous grants of time on the
National Energy Research Supercomputer Center.
The support of the U. S. Department of Energy is
gratefully acknowledged.

\newpage

\appendix
\section{Unitary transformation of the deuteron wave functions}
Consider the case of deuteron. The Hamiltonian is given by Eq. (\ref{H0})
for $\mu$=0. The wave function in momentum space is expressed as
\beq
|\psi_0({\bf p})\rangle = R_0(p){\cal Y}_{110}(\hat{\bf p}) +
R_2(p){\cal Y}_{112}(\hat{\bf p})
\eeq
where $R_0(p)$ and $R_2(p)$ are the $S$- and $D$-state radial wave functions,
${\cal Y}_{LSJ}(\hat{\bf p})$ are the spin-angle functions
\beq
{\cal Y}_{LSJ}=(-i)^L\sum_{M_S}\langle L,M_L=M_J-M_S,S,M_S|J,M_J\rangle
Y_{L,M_L}|S,M_S\rangle
\eeq
Here $J, S, L$ are the total, spin, and orbital angular momentum, respectively,
and $M_J,  M_S, M_L$ are their projections along $z$-axis.
From Eq. (\ref{unitary-wave}), the wave functions of $\mu$=$\pm 1$ are
given by
\beq
|\psi({\bf p})^{\mu=\pm 1}\rangle^{\rm uni} = |\psi_0({\bf p})\rangle
-\int\frac{d^3p}{(2\pi)^3}iU({\bf p},{\bf p}^{\prime})
|\psi_0({\bf p}^{\prime})\rangle
\eeq
where the unitary operator is given by Eq. (\ref{unitary-op}) multiplied by
the form factor [Eq. (\ref{ff})]. We can rewrite
\beq
\frac{1}{m_{\pi}^2+q^2}\left(\frac{\Lambda_{\pi}^2-m_{\pi}^2}{\Lambda_{\pi}^2
+q^2}\right)^2=\frac{1}{m_{\pi}^2+q^2}-\frac{1}{\Lambda_{\pi}^2+q^2}
+(\Lambda_{\pi}^2-m_{\pi}^2)\frac{d}{d\Lambda_{\pi}^2}
\frac{1}{\Lambda_{\pi}^2+q^2}.
\label{vertex}
\eeq
Then use
\beq
\frac{1}{m_{\pi}^2+q^2}=\frac{2\pi}{pp^{\prime}}\sum_{lm}Q_l(z_{m_{\pi}})
Y_{lm}^*(\hat{\bf p})Y_{lm}(\hat{\bf p}^{\prime})
\eeq
where $Q_l(z_{m_{\pi}})$ is the Legendre function of the second kind and
\beq
z_{m_{\pi}}=\frac{p^2+p^{\prime 2}+m_{\pi}^2}{2pp^{\prime}}
\eeq
Because deuteron has only $l$=0 and 2 states, we will only need $Q_0$ and $Q_2$
\ba
Q_0(z) &=& \frac{1}{2}\mbox{ln}\left(\frac{z+1}{z-1}\right), \nonumber \\
Q_2(z) &=& \frac{1}{2}(3z^2-1)Q_0(z)-\frac{3}{2}z
\ea
Using these, Eq. (\ref{vertex}) becomes
\beq
\frac{1}{m_{\pi}^2+q^2}\left(\frac{\Lambda_{\pi}^2-m_{\pi}^2}{\Lambda_{\pi}^2
+q^2}\right)^2 = \frac{2\pi}{pp^{\prime}}\sum_{lm}\bar{Q}_l(p^{\prime},p)
Y_{lm}^*(\hat{\bf p})Y_{lm}(\hat{\bf p}^{\prime})
\eeq
where
\beq
\bar{Q}_l(p^{\prime},p) = Q_l(z_{m_{\pi}})-Q_l(z_{\Lambda_{\pi}})
+(\Lambda_{\pi}^2-m_{\pi}^2)\frac{d}{d\Lambda_{\pi}^2}Q_l(z_{\Lambda_{\pi}})
\eeq
Then we use the following relations
\ba
\boldsigma_1\cdot{\bf p}\ \boldsigma_2\cdot{\bf p}&=&\frac{1}{3}p^2
\left(S_{12}(\hat{\bf p})+\boldsigma_1\cdot\boldsigma_2\right), \\
S_{12}(\hat{\bf p}){\cal Y}_{110}(\hat{\bf p}) &=& \sqrt{8}{\cal Y}_{112}
(\hat{\bf p}), \\
S_{12}(\hat{\bf p}){\cal Y}_{112}(\hat{\bf p}) &=& \sqrt{8}{\cal Y}_{110}
(\hat{\bf p})-2{\cal Y}_{112}(\hat{\bf p}), \\
\boldsigma_1\cdot\boldsigma_2 {\cal Y}_{110}(\hat{\bf p}) &=&
{\cal Y}_{110}(\hat{\bf p}), \\
\boldsigma_1\cdot\boldsigma_2 {\cal Y}_{112}(\hat{\bf p}) &=&
{\cal Y}_{112}(\hat{\bf p}).
\ea
Finally, putting everything together, we get
\beq
|\psi^{\mu}({\bf p})\rangle^{\rm uni} = R_0^{\rm uni}(p){\cal Y}_{110}
(\hat{\bf p}) + R_2^{\rm uni}(p){\cal Y}_{112}(\hat{\bf p})
\eeq
where
\ba
R_0^{\rm uni}(p) &=& R_0(p) + \mu\times\frac{1}{8\pi^2}\frac{f_{\pi NN}^2}
{m_{\pi}^2}\frac{m^2}{Ep}\int\frac{p^{\prime}dp^{\prime}}
{E^{\prime}}\left\{(E^{\prime}-E)\bar{Q}_0(p^{\prime},p)R_0(p^{\prime})\right.
\nonumber \\
& &\hspace{1.1cm}+\left.\sqrt{8}\left[(E^{\prime}-m)\bar{Q}_0(p^{\prime},p)
-(E-m)\bar{Q}_2(p^{\prime},p)\right]R_2(p^{\prime})\right\} \nonumber \\
R_2^{\rm uni}(p) &=& R_2(p) + \mu\times\frac{1}{8\pi^2}\frac{f_{\pi NN}^2}
{m_{\pi}^2}\frac{m^2}{Ep}\int\frac{p^{\prime}dp^{\prime}}
{E^{\prime}}\left\{\sqrt{8}(E^{\prime}-E)\bar{Q}_0(p^{\prime},p)R_0(p^{\prime})
\right. \nonumber \\
& &\hspace{1.1cm}+\left.\left[(E^{\prime}-m)\bar{Q}_0(p^{\prime},p)-(E-m)
\bar{Q}_2(p^{\prime},p)\right]R_2(p^{\prime})\right\}
\label{unitary-SDwave}
\ea
The wave functions should be normalized to 1, then Fourier transformed to
configuration space to obtain the $S$- and $D$-state wave functions shown
in Fig. \ref{fig:unitary}.

\newpage

\begin{figure}
\caption{Deuteron $S$- and $D$-state wave functions.}
\label{fig:wavefunc}
\end{figure}

\begin{figure}
\caption{Deuteron charge form factors squared for spin projections
$M_d$=0 and 1 in impulse approximation, for various potentials.}
\label{fig:fc}
\end{figure}

\begin{figure}
\caption{Deuteron tensor polarizing power $T_{20}$ in impulse approximation,
for various potentials.}
\label{fig:t20}
\end{figure}

\begin{figure}
\caption{Comparison of deuteron wave functions obtained from unitary
transformation and those from solving Schr\"odinger equation.}
\label{fig:unitary}
\end{figure}

\begin{figure}
\caption{Deuteron $D$-sate probability as a function of $\mu$. The circles
represent the $P_D$ obtained by solving Schr\"odinger equation, while
the diamond symbols represent those from unitary transformation.}
\label{fig:pd}
\end{figure}

\begin{figure}
\caption{Comparison of the ``local $\mu$=$-1$'' central and tensor
potentials with AV18. }
\label{fig:vct}
\end{figure}

\newpage

\begin{table}
\caption{Deuteron properties predicted by the modern potentials and
$v_{18}^{\mu}$ for $\mu$=$-$1, 0, +1.}
\vspace{0.5cm}
\begin{tabular}{lrrrr}
                 & Character &$Q_d$ (fm$^2$)& $D/S$ ratio & $P_D$ (\%) \\ \hline
Experiment       & nonlocal  & 0.2859(3)    & 0.0256(4)   & --         \\ \hline
AV18             & local     & 0.270        & 0.0250      & 5.76       \\
Nijmegen II      & local     & 0.271        & 0.0252      & 5.64       \\
Reid93           & local     & 0.270        & 0.0251      & 5.70       \\
Nijmegen I       & nonlocal  & 0.272        & 0.0253      & 5.66       \\
CD-Bonn ($\mu$=$-1$)& nonlocal& 0.270       & 0.0255      & 4.83       \\ \hline
$v_{18}^{\mu=-1}$& nonlocal  &  0.266       & 0.0253      & 4.98       \\
$v_{18}^{\mu=0}$ & nonlocal  &  0.271       & 0.0258      & 5.73       \\
$v_{18}^{\mu=+1}$& nonlocal  &  0.272       & 0.0260      & 6.26
\end{tabular}
\end{table}

\begin{table}
\caption{Expectation values of kinetic energy $T$, OPEP $v_{\pi}$ and
the phenomenological short-range part $v_R$ of various potentials (in MeV).}
\vspace{0.5cm}
\begin{tabular}{lrrrr}
            & AV18   & $v_{18}^{\mu=-1}$ & $v_{18}^{\mu=0}$ & $v_{18}^{\mu=+1}$
\\ \hline
$\langle E\rangle^\dagger$ & $-$2.242 & $-$2.242 & $-$2.242 & $-$2.242 \\ \hline
$\langle T\rangle$         &  19.882  &  17.352  &  18.877  &  20.161  \\ 
$\langle v_{\pi} \rangle$  &$-$21.355 &$-$15.642 &$-$18.797 &$-$21.486 \\
$\langle v_R \rangle$      & $-$0.769 & $-$3.952 & $-$2.322 & $-$0.917 \\ \hline
$\langle v_{\pi}^{\rm on-shell}\rangle$&$-$21.355&$-$17.399&$-$18.797 &$-$19.407  \\
$\langle v_{\pi}^{\rm off-shell}\rangle$& 0  &   1.757 &   0     & $-$2.079
\end{tabular}
$\dagger$ without electromagnetic interactions
\end{table}

\begin{table}
\caption{Deuteron $D$-state probability $P_D$, quadrupole moment $Q_d$ and
asymptotic $D/S$ ratio for various values of $\mu$.}
\vspace{0.5cm}
\begin{tabular}{lrrr}
$\mu$         & $P_D$ (\%) & $Q_d$ (fm$^2$) & $D/S$ ratio \\ \hline
0 (exact)     & 5.73        & 0.271         & 0.0258      \\
$-$1 (unitary)& 5.16        & 0.268         & 0.0253      \\
$-$1 (exact)  & 4.98        & 0.266         & 0.0253      \\
+1 (unitary)  & 6.34        & 0.273         & 0.0263      \\
+1 (exact)    & 6.26        & 0.273         & 0.0260
\end{tabular}
\end{table}

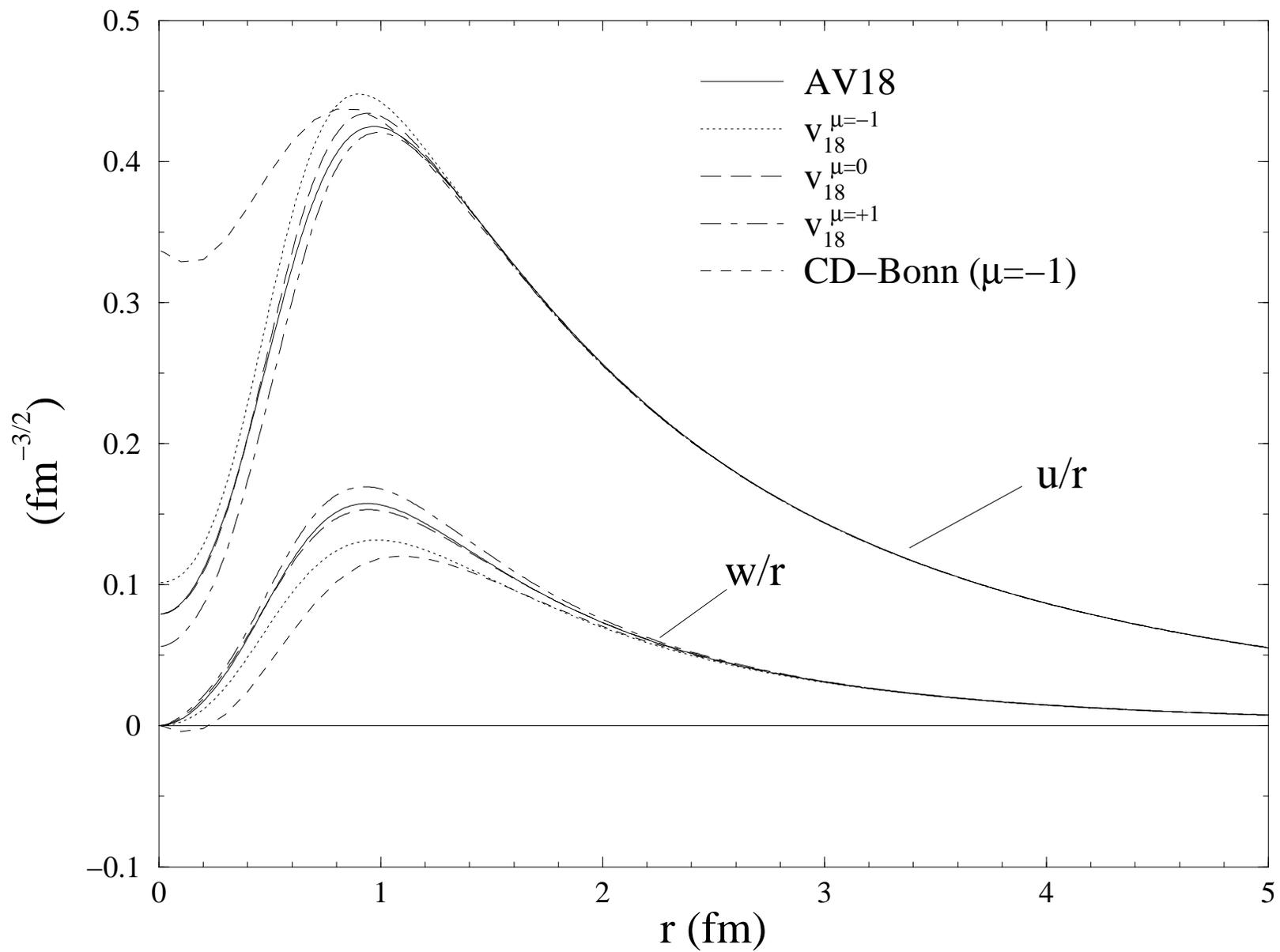
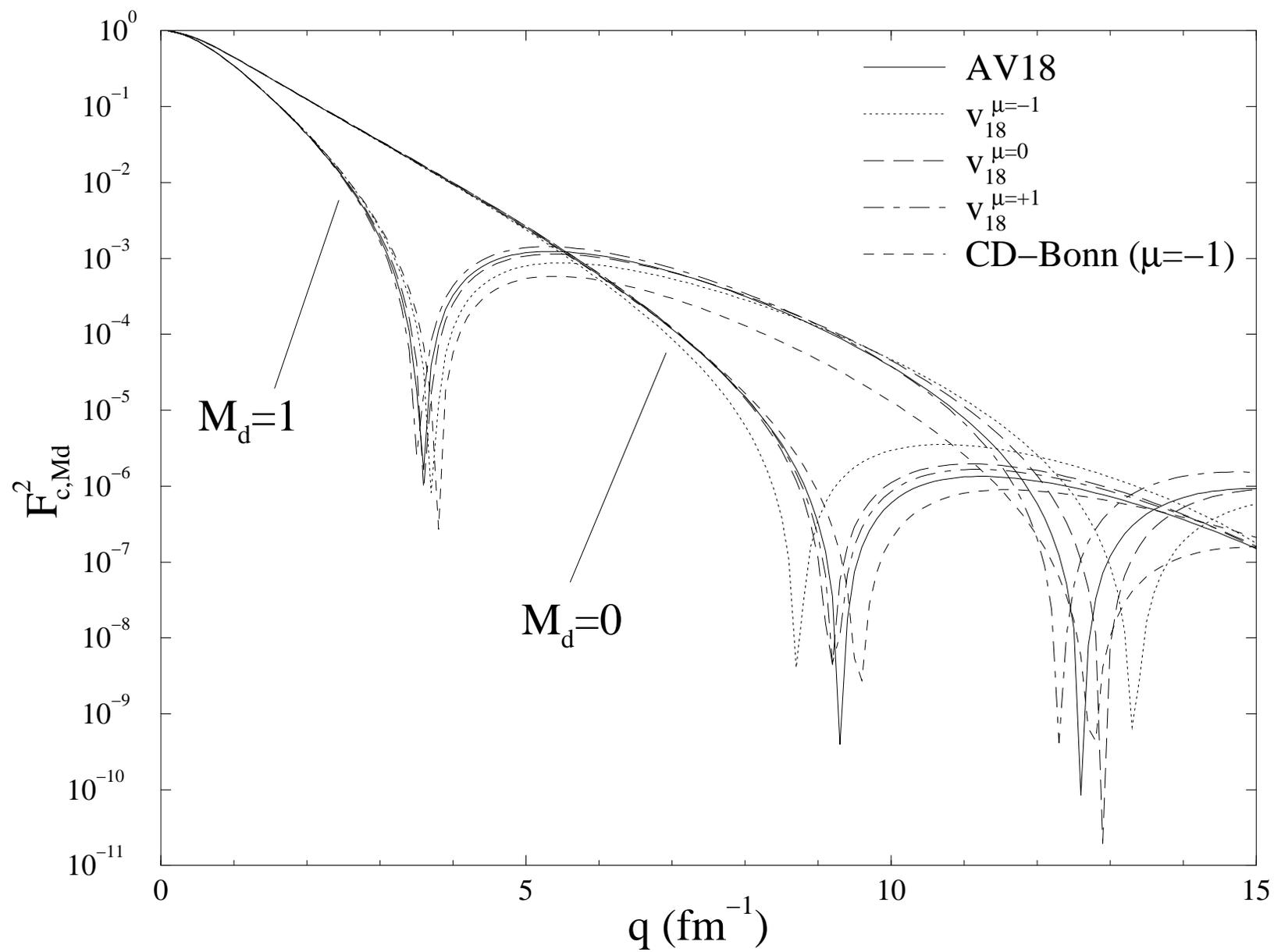
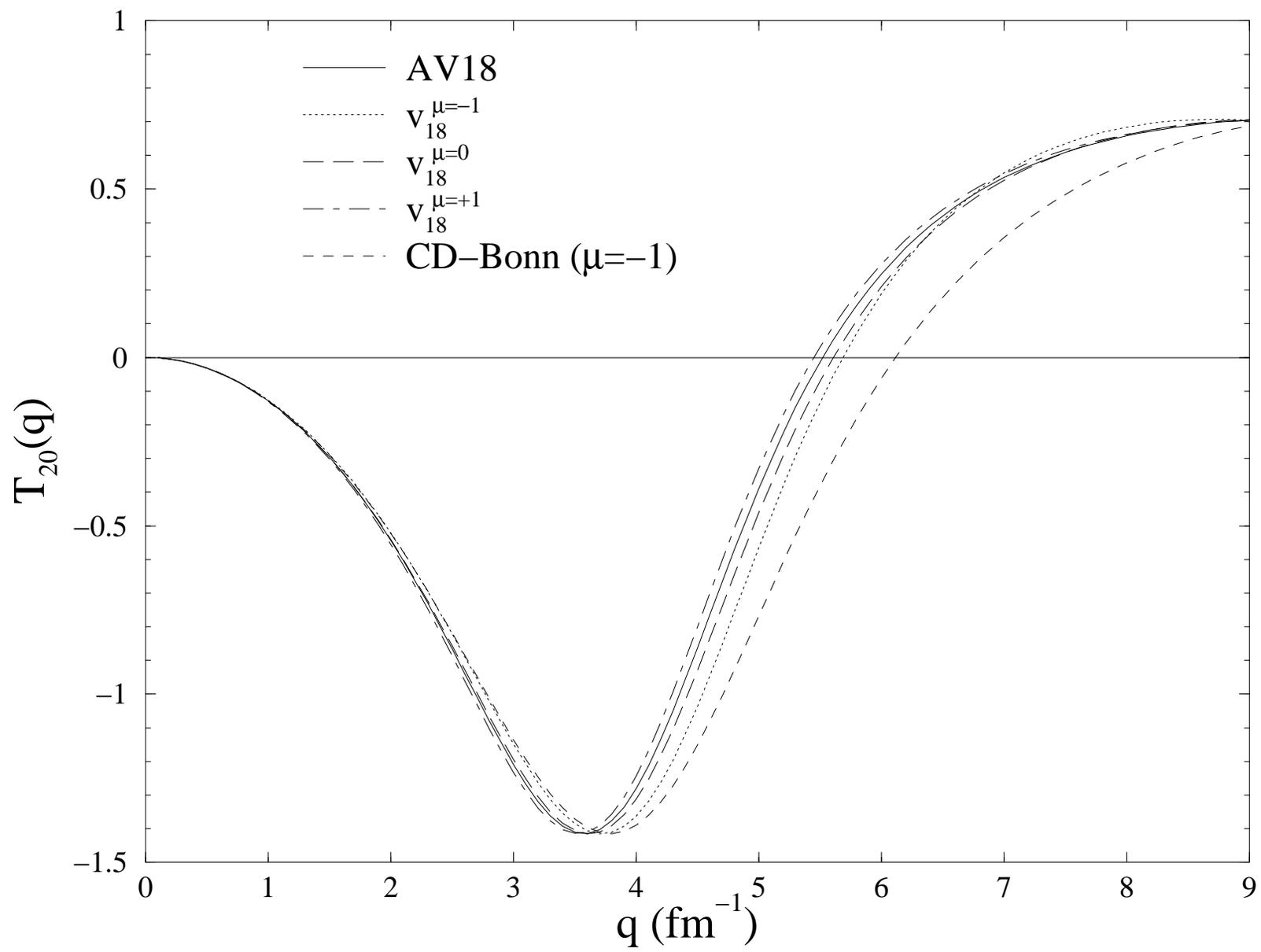
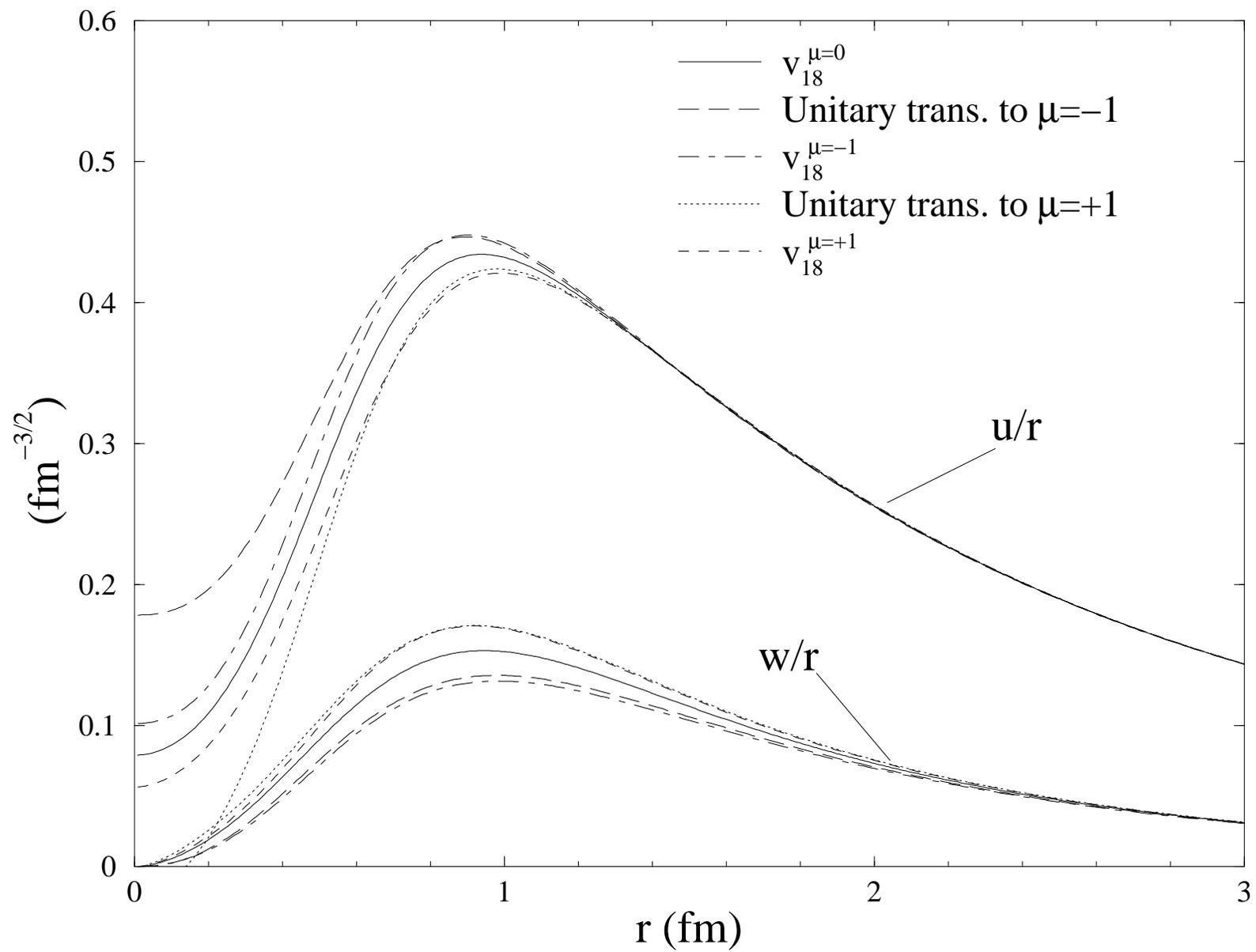
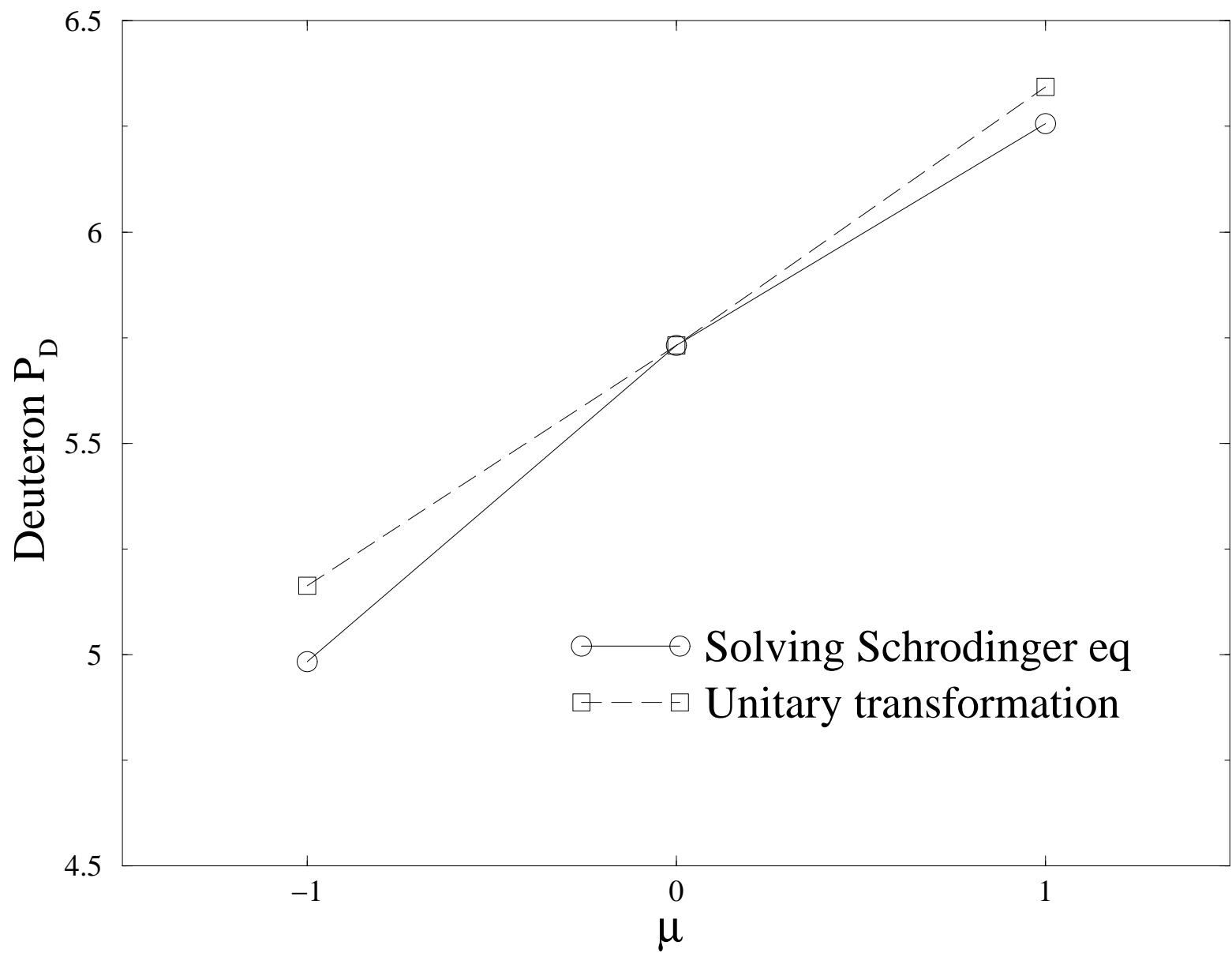
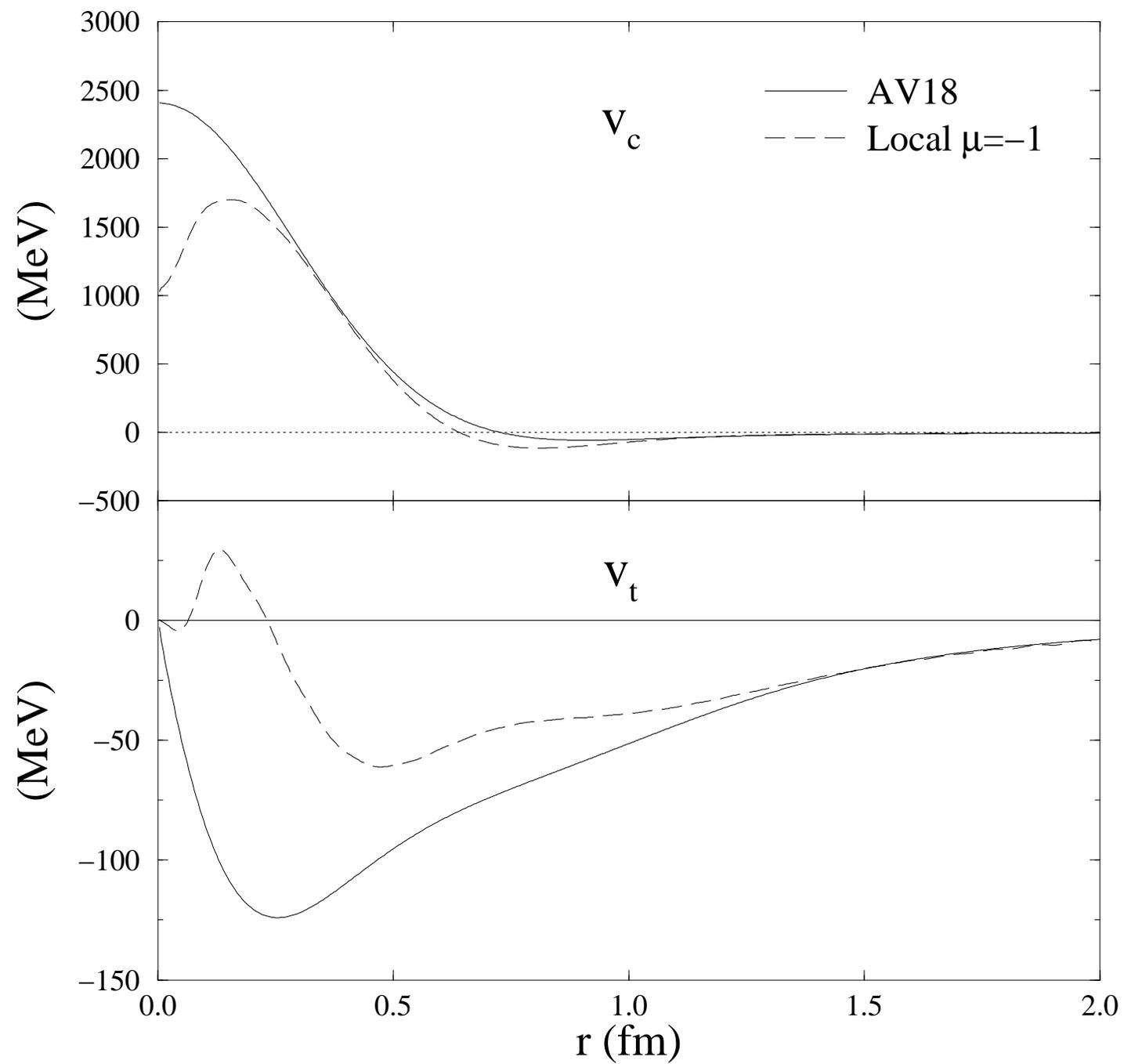
\end{document}